\documentclass[preprint2]{aastex}

\begin {document}

\title{A two-armed pattern in flickering maps of the nova-like 
       variable UU~Aquarii\thanks{Based on observations made at the 
       Laborat\'orio Nacional de Astrof\'{\i}sica, CNPq, Brazil.}}
\shorttitle{flickering maps of UU Aqr}

\author{Raymundo Baptista} 
\affil{Departamento de F\'{i}sica , Universidade Federal de Santa 
     Catarina, Campus Trindade, 88040-900, Florian\'opolis, SC, Brazil}
\email{bap@astro.ufsc.br}
\and 

\author{Alexandre Bortoletto}
\affil{Instituto de Astronomia, Geof\'isica e Ci\^encias Atmosf\'ericas, 
     Universidade de S\~ao Paulo, Rua do Mat\~ao 1228, 05508-900, 
     S\~ao Paulo, SP, Brazil}
\email{abortoletto@astro.iag.usp.br}

\begin{abstract} 
We report the analysis of a uniform sample of 31 light curves of 
the nova-like variable UU~Aqr with eclipse mapping techniques.
The data were combined to derive eclipse maps of the average steady-light 
component, the long-term brightness changes, and low- and high-frequency 
flickering components.  The long-term variability responsible for the 
'low' and 'high' brightness states is explained in terms of the response 
of a viscous disk to changes of 20-50 per cent in the mass transfer 
rate from the donor star.  
Low- and high-frequency flickering maps are dominated by emission from
two asymmetric arcs reminiscent of those seen in the outbursting dwarf 
nova IP~Peg, and are similarly interpreted as manifestation of a 
tidally-induced spiral shock wave in the outer regions of a large 
accretion disk.
The asymmetric arcs are also seen in the map of the steady-light
aside of the broad brightness distribution of a roughly steady-state 
disk. The arcs account for 25 per cent of the steady-light flux and
are a long-lasting feature in the accretion disk of UU~Aqr.
We infer an opening angle of $10\degr\pm3\degr$ for the spiral arcs. 
The results suggest that the flickering in UU~Aqr is caused by 
turbulence generated after the collision of disk gas with the 
density-enhanced spiral wave in the accretion disk.
\end{abstract}


\section{Introduction}

In Cataclysmic Variables (CVs) a late-type star overfills its Roche 
lobe and transfers matter to a companion white dwarf (WD) via an 
accretion column or disk. The CV zoo comprises low-mass transfer dwarf 
novae and high-mass transfer novae and nova-like systems. The light 
curve of these binaries show intrinsic brightness fluctuations 
(flickering) of 0.1-1 mag on timescales from seconds to tens of minutes, 
considered a basic signature of accretion (Warner 1995). 
When the Roche lobe of the WD is suficiently large there is room for 
the accretion disk to expand beyond the 3:1 resonance radius and the
tidal pull of the mass-donor star becomes relevant to the gas dynamics, 
giving rise to precessing elliptical rings (when the binary mass ratio 
$q=M_2/M_1 < q_\mathrm{crit}$, e.g. Whitehurst \& King 1991) and 
spiral shock waves (e.g., Steeghs 2001) in the outer disk regions.  
While theoretical expectations and numerical simulations suggest 
$q_\mathrm{crit}\simeq 0.3$ (e.g., Osaki 1996; Kunze, Speith \& Riffert
1997; Murray, Warner \& Wickramasinghe 2000), an extensive study of
superhumps allowed Patterson et~al. (2005) to set an observational upper 
limit of $q_\mathrm{crit}=0.35 \pm 0.02$ for the creation of superhumps.

Earlier studies (Bruch 1992, 1996, 2000) led to the suggestion that 
there are mainly two sources of flickering in CVs, the stream-disk 
impact region at disk rim and a turbulent inner disk region in the 
vinicity of the WD (possibly the boundary layer), the relative 
importance of which varies from system to system. 
The spatially-resolved study of flickering in the dwarf nova 
V2051~Oph by Baptista \& Bortoletto (2004, hereafter BB04) revealed a 
more complex scenario, in which the low-frequency flickering is 
associated to an overflowing gas stream (possibly as a consequence of
unsteady mass transfer from the mass-donor star, e.g., Warner \& Nather 
1971) and the high-frequency flickering is distributed over the surface
of the accretion disk -- possibly as a consequence of magnetohydrodynamic 
(MHD) turbulence or events of magnetic reconnection at the disk 
chromosphere (Geertsema \& Achterberg 1992; Kawagushi et~al. 2000). 
The identification of a disk component to the flickering and the
consequent estimation of the disk viscosity $\alpha$-parameter (Shakura
\& Sunyaev 1973) through the application of an MHD turbulence model, 
raised the expectation that the technique could be applied to measure
the accretion disk viscosity of other CVs. In particular, the tight 
correlation between the flickering disk component and the steady-light
emission in V2051~Oph (BB04) suggested that this could be the 
dominant source of flickering in nova-like systems, with their hot and 
bright accretion disks. An obvious next step would be to perform a 
spatially-resolved study of flickering on an eclipsing nova-like system
to test these ideas.

UU~Aqr is a bright eclipsing nova-like variable with an orbital period 
of 3.9~hr and a mass ratio $q=0.30\pm 0.07$ (Baptista, Steiner \& 
Cieslinski 1994; hereafter BSC). It shows conspicuous flickering 
activity (with flares which may reach 25\% of the total system light), 
long-term ($\sim 4\;yr$) changes of 0.3~mag in brightness (atributted 
to variations in the mass transfer rate from the mass-donor star, BSC), 
as well as $\sim 1$~mag changes on timescales of days (interpreted 
as being 'stunted' dwarf nova-type outbursts, Honeycutt, Robertson \& 
Turner 1998). It also displays superhumps in its light curve (Patterson 
et~al. 2005), suggesting the presence of a large, elliptical precessing 
accretion disk.

Here we report the analysis of a large sample of light curves of UU~Aqr
with eclipse mapping techniques to locate the sources of the flickering 
and to investigate the long-term brightness changes in this binary.

\section{Observations} \label{observa}

Time series of $B$-band CCD photometry of UU~Aqr were obtained with 
an EEV camera ($385\times 578$~pixels, $0.58\arcsec {\rm pixel}^{-1}$)
attached to the 0.6~m telescopes of the Laborat\'orio Nacional de
Astrof\'{\i}sica, in southern Brazil, from 1998 to 2001.  The CCD 
camera is operated in frame transfer mode, with negligible (13~ms) 
dead time between exposures. It has a Global Positioning System board 
that sets its internal clock to UTC to an accuracy better than 10~ms.  
The observations are summarized in Table~\ref{tab1}.
Columns 2 and 8 list the number of points in the light curve ($N_p$),
columns 3 and 9 give the exposure time in seconds ($\Delta t$). 
Columns 4 and 10 list the eclipse cycle number (E); the numbers in 
parenthesis indicate observations that, because of incomplete phase 
coverage or interruptions caused by clouds, do not cover the eclipse 
itself. Columns 6 and 12 give an estimate of the quality of the 
observations.  The seeing ranged from $1.5\arcsec$ to $2.5\arcsec$.
The observations comprise 31 light curves obtained with the same 
instrument and telescope, which ensures a high degree of uniformity 
to the data set. 
%
%%%%%%%%%%%%%%%%%%%%%%%%%%%%%%%%%%%%%%%%%%%%% TABLE 1
\placetable{tab1}

Data reduction procedures included bias subtraction, flat-field 
correction, cosmic rays removal and aperture photometry extraction. 
Time series were constructed by computing the magnitude difference 
between the variable and a bright reference comparison star 
$77\arcsec$~E and $47\arcsec$~S of the variable with scripts based on 
the aperture photometry routines of the APPHOT/IRAF package
 \footnote{IRAF is distributed by National Optical Astronomy 
  Observatories, which is operated by the Association of Universities 
  for Research in Astronomy, Inc., under contract with the National 
  Science Foundation.}.  
Light curves of other comparison stars in the field were also computed 
in order to check the quality of the night and the internal consistency
and stability of the photometry over the time span of the observations.  
The magnitude and colors of the reference star were tied to the 
Johnsons-Cousins UBVRI systems (Bessell 1990) from observations of this 
star and of standard stars (Graham 1982; Stone \& Baldwin 1983) made 
on 4 photometric nights. We used the relations of Lamla (1981) to 
transform UBVRI magnitudes to flux units.  The B-band flux of the 
reference star was then used to transform the light curves of the 
variable from magnitude difference to absolute flux. We estimate that 
the absolute photometric accuracy of these observations is about 12 
per cent. On the other hand, the analysis of the light curves of field
comparison stars of brightness comparable to that of the variable 
indicates that the internal error of the photometry is less than 2 
per cent. The error in the photometry of the variable is derived from 
the photon count noise and is transformed to flux units using the 
same relation applied to the data. The individual light curves have 
typical signal-to-noise ratios of $S/N=$40-60 out-of-eclipse and 
$S/N=$10-20 at mid-eclipse.

\section{Data analysis} \label{analysis}

The light curves were phase-folded according to the linear ephemeris
(Borges 2005, private communication),
\begin{equation}
T_{mid}=HJD\; 2\,446\,347.2659 + 0.163\,804\,9430 \times E \; ,
\label{efem}
\end{equation}  
where $T_{mid}$ gives the WD mid-eclipse times.
Fig.\,\ref{fig1} shows the light curves of UU Aqr superimposed in 
phase. The upper frame depicts the light curves of a comparison star 
with the same brightness of UU~Aqr around mid-eclipse. The constancy 
of the comparison star flux over time indicates that all brightness 
variations seen in UU~Aqr are intrinsic to the variable. The scatter 
with respect to the mean level is significantly larger in UU Aqr and is 
caused by a combination of flickering and long term brightness changes. 
\placefigure{fig1}

We applied the `single' (Bruch 1996, 2000) and `ensemble' (Horne \& 
Stiening 1985; Bennie, Hilditch \& Horne 1996) methods to the set of
light curves of UU~Aqr to derive the orbital dependency of its 
steady-light, long-term brightness changes, low- and high-frequency
flickering components. The reader is referred to BB04 for a detailed
description and combined application of both methods.

UU~Aqr was in its 'high' brightness state during the 1998 and 2001
observations, and in its 'low' brightness state during the 1999 and 
2000 runs. The curves of the 'high' and 'low' brightness states are 
identified in Fig.~\ref{fig1} by black and grey symbols, respectively. 
We remark that the average out-of-eclipse flux level increases steadily 
from the 1999 to the 2001 data and that the nominal separation between
the 'low' and 'high' brightness states is a rather arbitrary one. 
Because of the scatter produced by the strong flickering (with an 
average pear-to-peak amplitude of $\simeq 3$~mJy for both brightness
levels), there is an overlap in flux between the curves of the 'low'
and 'high' states.  In order to test the influence of the brightness 
state on the flickering behaviour, we applied the 'ensemble' method 
separately for the data of the 'high' and 'low' brightness states. 
We found no evidence of a dependency of the flux level or eclipse shape
of the flickering curve with brightness level in UU~Aqr. We therefore
combined all light curves for the following analysis. The difference 
in average brightness level seen in UU~Aqr along the observations is 
properly taken into account in the ensemble method by the curve of 
the long-term changes (see below, BB04).

In order to apply the 'ensemble' method, we define a reference 
out-of-eclipse flux $f_{ref}$ (the mean flux over the phase range
$0.15-0.80$) for each individual light curve and we divide the data
into a set of phase bins. A linear fit to the $f_i \times f_{ref}(i)$ 
diagram for each phase bin (e.g., Fig.~2 of BB04) yields an 
average flux (the steady-light component), an angular coefficient 
(which measures the long-term change) and a standard deviation with 
respect to the linear fit (the scatter curve, with added contributions 
from the Poisson noise and the flickering). 
We multiply the non-dimensional angular coefficients by the amplitude 
of the variation of the reference flux in the data set, 
$\Delta f_{ref}= 5.5\pm0.1$~mJy, to express the long-term changes in 
terms of the amplitude of the flux change per phase bin, 
$\Delta f_\nu(\phi)$. 

The average steady-light curve was subtracted from each individual 
light curve to remove the DC component and a Lomb-Scargle periodgram 
(Press et~al. 1992) was calculated. The periodgrams of all light curves 
were combined to yield a mean periodgram and a standard deviation with 
respect to the mean.  Fig.~\ref{fig2} shows the resulting average power 
density spectrum (PDS) binned to a resolution of 0.02 units in 
log(frequency). The PDS is well described by a power-law $P(f) \propto 
f^{-1.5}$. It becomes flat for $f_{low} < 0.15\;mHz \; (t_{low} > 111$ 
minutes) and disappears in the white noise for $f_{up}> 20\;mHz 
\;(t_{up}< 50\,s$). The slope of the PDS distribution is reminiscent
of those seen in other CVs, which can be well described by power laws 
with an average exponent $P(f) \propto f^{-2.0\pm0.8}$ (Bruch 1992).
\placefigure{fig2}

The 'ensemble' method samples flickering at all frequencies. But 
because of the power-law dependency of the flickering, an 'ensemble' 
curve is dominated by the low-frequency flickering components. 
On the other hand, the filtering process of the `single' method 
produces curves which sample the high-frequency flickering. The 
combination of both methods allows one to separate the low- (ensemble)
and high-frequency (single) components of the flickering.

The Achilles heel of the 'single' method is its difficulty in 
separating the high-frequency flickering from the rapid brightness
changes caused by the eclipse. 
% This problem imposes restrictions in the range of smoothing filter's 
% cutoff frequency applied in the method. 
In order to overcome this limitation, we subtracted the average 
steady-light curve from each individual light curve before applying the 
'single' filtering process to eliminate the steep gradients produced by 
the eclipse in the light curve.
Our 'single' light curve includes flickering components with frequencies
$f_c > 2\;mHz$ (timescales shorter than $t_c=500\,s$, Fig.~\ref{fig2}). 
Single curves obtained with cut-off frequencies of $f_c= 3\;mHz\;
(t_c= 333\,s)$ and $f_c= 5\;mHz\;(t_c= 100\,s)$ show the same 
morphology of the lower cutoff frequency curve.  Because of the reduced
power, these other curves are noisier and will not be presented here.

The steady-light, long-term changes, and flickering curves were
analyzed with eclipse mapping techniques (Horne 1985; Baptista \& 
Steiner 1993) to solve for a map of the disk surface brightness 
distribution and for the flux of an additional uneclipsed component
in each case.  The uneclipsed component accounts for all light that is
not contained in the eclipse map (i.e., light from the secondary star
and/or a vertically extended disk wind). 
The reader is referred to Rutten, van Paradijs \& Tinbergen (1992) 
and Baptista, Steiner \& Horne (1996, hereafter BSH) for a detailed 
description of and tests with the uneclipsed component, and to 
Baptista (2001) for a recent review on the eclipse mapping method.
Out-of-eclipse brightness changes (not accounted for by the standard
eclipse mapping method) were removed from the light curves by fitting
a spline function to the phases outside eclipse, dividing the light 
curve by the fitted spline, and scaling the result to the spline 
function value at phase zero (e.g., BSH). 

Our eclipse map is a flat Cartesian grid of $51\times51$ pixels 
centered on the white dwarf with side $2 R_{L1}$ (where $R_{L1}$ is 
the distance from the disk center to the inner Lagrangian point L1).
The eclipse geometry is defined by the mass ratio $q$ and the 
inclination $i$, and the scale of the map is set by $R_{L1}$. We 
adopted $R_{L1}=0.744\;R_\odot$, $q=0.3$ and $i=78\degr$ (BSC), which
correspond to a white dwarf eclipse width of $\Delta\phi = 0.051$~cycle.
This combination of parameters ensures that the white dwarf is at the
center of the map.  The reconstructions were performed with a polar 
gaussian default function (Rutten et~al. 1992) with radial blur width 
$\Delta r= 0.02\;R_{L1}$ and azimuthal blur width $\Delta\theta= 
30\degr$, and reached a final reduced chi-square $\chi_\nu^{2}\simeq 1$
for all light curves.  The uncertainties in the eclipse maps were 
derived from Monte Carlo simulations with the light curves using a 
bootstrap method (Efron 1982, Watson \& Dhillon 2001), generating a
set of 20 randomized eclipse maps (see Rutten et~al. 1992). These are
combined to produce a map of the standard deviations with respect to
the true map. A map of the statistical significance (or the inverse of
the relative error) is obtained by dividing the true eclipse map by 
the map of the standard deviations (Baptista et~al. 2005).

\section{Results}

Light curves and corresponding eclipse maps are shown in 
Fig.~\ref{fig3}.  For a better visualization of structures in the disk
brightness distributions, the asymmetric disk components are also 
shown. An asymmetric component is obtained by slicing the disk into a 
set of radial bins and fitting a smooth spline function to the mean 
of the lower half of the intensities in each bin. The spline-fitted 
intensity in each annular section is taken as the symmetric 
disk-emission component. The asymmetric disk component is then 
obtained by subtracting the symmetric disk from the original eclipse 
map (eg., Saito \& Baptista 2006). This procedure removes the baseline
of the radial distribution while preserving all azimuthal structure. 
\placefigure{fig3}

\subsection{Steady-light and long-term changes} \label{steady}

The steady-light light curve gives the flux per phase bin for the 
mid-reference flux level and represents the median steady brightness
level along the data set. Because it is obtained by combining 31 
light curves, it has high S/N and the corresponding eclipse map is
of high statistical significance (typically $>10\,\sigma$).

The eclipse map of the steady-light shows an extended brightness 
distribution peaking at disk center with two asymmetric arcs on
roughly opposite disk sides (Fig.~\ref{fig3}, top row).
The asymmetries are diluted by the dominant broad disk brightness 
distribution and only become clear in the asymmetric disk component. 
The arcs account for $\simeq 25$ per cent of the total flux of the 
steady light map. They are located at different radii, with the one 
in the trailing side (the lower disk hemisphere in the eclipse maps 
of Fig.~\ref{fig3}) being closer to disk center.  
The asymmetric arcs do not coincide with the WD at disk center or 
bright spot at disk rim. 

By transforming the intensities in the steady-light eclipse map into 
blackbody brightness temperatures (assuming a distance of 200\,pc to 
the binary, BSH) we find that the radial temperature distribution 
closely follows the $T\propto R^{-3/4}$ law for steady accretion in 
the outer disk and becomes flatter in the inner disk regions ($R 
\leq 0.2\,R_{L1}$), leading to estimated mass accretion rates of 
\.{M}$=10^{-9.0\pm0.2}\;M_\odot\;yr^{-1}$ at $R=0.1\,R_{L1}$ and 
\.{M}$=10^{-8.80\pm0.06}\;M_\odot\;yr^{-1}$ at $R=0.3\,R_{L1}$.
The brightness temperatures decrease steadily with radius, from 
$\simeq 13000\,K$ at $0.1\,R_{L1}$, to $\simeq 9400-7000\,K$ at
$(0.2-0.4)\,R_{L1}$ (the radial range at which the asymmetric arcs 
are located), and $\simeq 5500\,K$ at $0.5\,R_{L1}$.
We also find an uneclipsed component of $6.4 \pm 0.3$ per cent of 
the total steady-light flux.
The inferred brightness temperatures, uneclipsed component and mass
accretion rates are in good agreement with previous results (BSH, 
Baptista et~al. 2000; Vrielmann \& Baptista 2000). 

The curve of the long-term changes measures brightness changes on 
timescales longer than the orbital period. It allows us to visualize 
the differences in disk structure between the observed 'low' and 
'high' brightness states of UU~Aqr.

The light curve of the long-term changes shows an eclipse with a 
pronouced shoulder at egress phases. The resulting eclipse map 
(Fig.~\ref{fig3}, second row from top) has a bright source at disk 
center and an azimuthally extended ($\Delta\theta\simeq 90\degr$) 
bright spot at disk rim, similar to those found in the eclipse maps 
of the 'high' state of BSH. The uneclipsed component is negligible.
This map tells us that the difference between the 'low' and 'high' 
brightness states is caused by an increase in the luminosity of the 
outer parts of the disk (as previously found by BSH) but also by a 
comparable increase in brightness of the innermost disk regions.

For a fixed distance, the intensities in the eclipse map scale 
linearly with the flux in the light curve (see, e.g., Baptista 2001).
Because we choose to express the long-term changes curve in terms of 
the amplitude of the flux variation (Sect.~\ref{analysis}), the 
intensities in the corresponding eclipse map are given in terms of the
amplitude of the variation in intensity between the minimum and maximum
brightness states in the data set, $\Delta I_j$ (where $j$ refers to
each pixel in the eclipse map). Thus, eclipse maps representing the
minimum and maximum brightness distributions can be obtained by 
adding/subtracting the appropriate proportion of the long-term 
changes map to/from the steady-light map, $\bar{I_j}$,
\begin{mathletters}
\begin{equation}
I_j(\mathrm{max})= \bar{I_j} + \frac{1}{2} \Delta I_j \,\,\, , 
\end{equation}
\begin{equation}
I_j(\mathrm{min})= \bar{I_j} - \frac{1}{2} \Delta I_j \,\,\, .
\end{equation}
\end{mathletters}
As expected, the resulting minimum and maximum brightness maps 
(Fig.~\ref{fig4}) are similar to the $B$-band eclipse maps of the 
'low' and 'high' brightness states of BSH (see their Fig.\,3).
\placefigure{fig4}

We now turn our attention to the interpretation of the structures 
seen in the long-term changes map.
BSH suggested that the azimuthally extended spot seen in the 'high'
brightness state reflects long-term changes in luminosity caused by 
variations in mass input rate at the outer disk. 
Baptista et~al. (2000) noted that it could alternatively be the
signature of an elliptical (precessing) disk similar to those 
possibly present in SU~UMa stars during super-outbursts.

In order to test the second possibility, we searched for the presence
of superhumps in our data set. We combined the light curves for a 
given year of observations and computed Lomb-Scargle periodograms
after removing the eclipses from the data. We find no statistically
significant periodicities in the combined light curves at the range
of frequencies near the orbital frequency. The only exception is
the data of 1999 (JD 2,451,403-2,451,404, when UU~Aqr was on its 
lowest brightness level along our data set), which presents a clear 
signal at the orbital frequency. A sine fit to the data shows that
the peak occurs at phases $\phi_{max}=0.87-0.92$~cycle on all orbits, 
indicating that we are seeing the orbital hump and not a superhump.
We remark that, because our observations focus on the eclipse and 
frame relatively short time intervals (of at most 2-3 consecutive 
orbits) on each night, they are not as sensitive to the presence of 
superhumps as the much more intensive and well sampled data of the
campaign by Patterson et~al. (2005).
Therefore, the lack of positive identification of superhumps in
our data indicates that, if present, the superhump signal was
of low amplitude/intensity and would hardly be able to account for 
the extended bright spot seen in the long-term changes map.
We are thus left with the explanation of BSH.

Can the bright source at inner disk also be explained in terms of
changes in mass input rate?  In order to answer this question, we took 
a reference steady-state disk model with mass accretion rate 
$\hbox{\.{M}}_\mathrm{ref}$ and computed the difference between the 
(blackbody) brightness distributions of a test steady-state model with 
$\hbox{\.{M}}_\mathrm{test}= \beta\, \hbox{\.{M}}_\mathrm{ref}$ 
($\beta>1$) and the reference model. For an opaque steady-state disk, 
the change in effective temperature, $\partial T$, caused by a change 
in mass accretion rate, $\partial$\.{M}, has the same radial dependency
of the effective temperature (see, e.g., Frank, King \& Raine 1992),
\begin{equation}
\frac{\partial T}{\partial\hbox{\.{M}}} \propto R^{-3/4} 
\left[ 1 - \left( \frac{R_{wd}}{R} \right)^{1/2} \right]^{1/4} \,\, ,
\end{equation}
where $R_{wd}$ is the WD radius.
Thus, the difference in temperature (and the corresponding difference in 
blackbody intensity) between two steady-state disk models increases with 
decreasing radius and peaks near disk center (at $R= 49/36\,R_{wd}$). 
The difference in intensity also scales with ${\partial\hbox{\.{M}}}$.
In searching for the pair of ($\hbox{\.{M}}_\mathrm{ref}, \beta$) values 
of best fit to the observed brightness distribution we find 
$\hbox{\.{M}}_\mathrm{ref}= 10^{-9.1\pm 0.1}\;M_\odot\;yr^{-1}$ and 
$\beta= 1.35\pm 0.15$.
Fig.~\ref{fig5} shows that the radial distribution of the central source
of the long-term changes map is consistent, within the uncertainties, 
with the difference in intensity expected for an increase by 20-50 per 
cent in mass accretion rate of a steady-state disk with $10^{-9.1}\;
M_\odot\;yr^{-1}$. Because the radial temperature distribution of the
steady-light map is actually flatter than the $T \propto R^{-3/4}$ law
of steady-state disks, these results should be considered illustrative.
Nevertheless, the inferred range of mass accretion rates ($10^{-9.1} -
10^{-8.9}\,M_\odot\;yr^{-1}$) is in line with the values for the 'low'
and 'high' brightness states found by BSH.
\placefigure{fig5}

Because in a steady-state disk the mass accretion rate reflects the
mass transfer rate, $\hbox{\.{M}}_2$, we may interpret the map of the 
long-term changes in terms of the response of a high-viscosity disk to 
changes in $\hbox{\.{M}}_2$ of about 20-50 per cent.  
When $\hbox{\.{M}}_2$ increases, the luminosity of the bright spot at
disk rim increases, as well as that of the inner disk regions -- as a
consequence of the increase of mass accretion through a disk close to
a steady-state.

\subsection{low- and high-frequency flickering} \label{flick}

The 'ensemble' and 'single' curves show a double-stepped eclipse 
reminiscent of the occultation of the two-armed spiral structure seen 
in eclipse maps of the dwarf nova IP~Peg in outburst (e.g., Baptista,
Haswell \& Thomas 2002; Baptista et~al. 2005), and lead to similar 
two-armed asymmetric brightness distributions (Fig.~\ref{fig3}, the two 
lowermost rows). The solid contour line overploted on each eclipse 
map depicts the 3-$\sigma$ confidence level region as derived from the 
map of statistical significance in each case (Sect.~\ref{analysis}).
The asymmetric arcs are at or above the 3-$\sigma$ confidence level in 
both maps. Vertical tick marks in the 'ensemble' panel of Fig.~\ref{fig3} 
indicate the eclipse ingress/egress phases of the two bright arcs 
(labeled '1' and '2').  Their location is depicted in the asymmetric 
component of the ensemble map. A simple comparison reveals that these are 
the same asymmetric arcs seen in the steady-light map. The major 
difference is that the arcs dominate the emission in the flickering maps 
(the asymmetric component account for 53 and 41 per cent of the total 
flux, respectively for the 'ensemble' and 'single' maps).
As already noted, they do not coincide with the WD at disk center nor 
with the bright spot at disk rim.

Although it is possible to center the eclipse of source 1 by applying 
a phase shift to all light curves, this would lead to physically 
unplausible brightness distributions for the steady-light and long-term
changes maps -- with highly asymmetric brightness distributions where 
the main sources fall at positions which cannot be associated with 
either the WD, the bright spot or gas stream (e.g., the azimuthally
extended spot in the long-term changes map would fall at the edge of the
primary Roche lobe, far away from the gas stream trajectory). 
It is also not possible to interpret the observed asymmetries in terms 
of enhanced emission from an elliptical outer disk ring because the 
asymmetries lie well inside the disk, far from its edge.
Given the similarities with the IP~Peg eclipse maps (e.g., see Fig.~3
of Baptista et~al. 2002) and the lack of plausible alternative 
explanations, we interpret the asymmetries in the flickering and 
steady-light maps as consequence of tidally-induced spiral shock waves 
in the accretion disk of UU~Aqr (e.g., Sawada et~al. 1986).

Arc 1 is in the trailing side of the disk at $R_{s1}= 0.20\pm 0.05\;
R_{L1}$; Arc 2 is in the leading side of the disk and is farther away 
from disk center, $R_{s2}= 0.32 \pm 0.05\;R_{L1}$.
The two arcs have azimuthal extent $\Delta\theta \simeq 110 \degr$ 
and radial extent $\Delta R \simeq 0.2\,R_{L1}$.
Baptista et~al. (2005) devised a way to estimate the opening angle of 
the spirals from the azimuthal intensity distribution of the eclipse
maps. We applied the same method to the flickering maps to estimate an 
opening angle of $\phi=10\degr \pm 3\degr$, indicating that the spiral 
arms in UU~Aqr are more tightly wound than in IP~Peg at outburst 
($\phi= 14\degr - 34\degr$, Baptista et~al. 2005). They are also
systematically closer to disk center than the arms seen in IP~Peg (at 
average distances of 0.30 and $0.55\,R_{L1}$, see Baptista et~al. 2005).
Because the opening angle of the spiral arms scales with the disk 
temperature (e.g., Steeghs \& Stehle 1999), this suggests that the 
outer accretion disk of UU~Aqr is cooler than that of IP~Peg in 
outburst.

We find uneclipsed components of $13\pm3$ and $17\pm7$ per cent of 
the total flux, respectively for the 'ensemble' and 'single' maps. 
This suggests that a sizeable part of the flickering may arise from 
outside the orbital plane, perhaps in a vertically-extended disk 
chromosphere + wind.

The 'ensemble' map samples flickering at all frequencies while the
'single' map contains the high-frequency (timescales $<500\,s$) 
flickering components. It is possible to separate the contribution of
the low-frequency flickering by subtracting the 'single' map from the 
'ensemble' map.  

Fig.~\ref{fig6} compares the relative amplitude of the low- ('ensemble' 
- 'single') and high-frequency ('single') flickering components in UU~Aqr. 
The radial run of the relative amplitudes are obtained by dividing the 
average radial intensity distribution of these two flickering maps by 
that of the steady-light. Dashed and dotted lines show the 1-$\sigma$ 
limits on the average amplitude, respectively for the low- and 
high-frequency flickering. The large uncertainties reflect the scatter 
introduced by the asymmetric arcs in the radial bins.
The two distributions are comparable, within the uncertainties.
Flickering is negligible in the inner disk. The amplitude of the 
low-frequency flickering increases monotonically with radius, reaching 
6 per cent of the total light at $0.3\,R_{L1}$. The amplitude of the 
high-frequency flickering also increases with radius and peaks at the 
location of the spiral arms ($\simeq 0.3\,R_{L1}$). 
The distributions are not reliable for $R\geq 0.45\,R_{L1}$ because
of the reduced statistical significance of the flickering maps and the
rapidly declining intensities in the steady-light map.
The lower panel of Fig.~\ref{fig6} depicts the ratio of the 'single'
to the 'ensemble' distributions and measures the contribution of the
high-frequencies ('single') to the total flickering.
The high-frequency flickering accounts for a roughly constant fraction
of $\simeq 50$ per cent of the flickering signal at all radii. 
There is no apparent difference in radial behaviour between the low- 
and high-frequency flickering components, indicating that they not only 
arise from the same location but are also produced by the same physical
mechanism.

We further notice that our data cover a time interval of about 4~yr
and that the eclipse maps yield the average behaviour over this time
scale -- with the implication that the observed spiral shocks are a 
long-lasting feature in the accretion disk of UU~Aqr.

\section{Discussion}

BB04 found two independent sources of flickering in their study of
the dwarf nova V2051~Oph: the low-frequency flickering is associated
to inhomogeneities in the mass transfer process, while the high-frequency
flickering originates in the accretion disk, possibly as a consequence
of MHD turbulence (Geertsema \& Achterberg 1992) or events of magnetic 
reconnection in the disk surface (Kawagushi et~al. 2000).

Contrary to the suggestion of Bruch (2000), UU~Aqr shows no evidence of 
flickering arising from a turbulent inner disk or from the bright spot. 
And, in contrast to V2051~Oph, it shows no disk-related flickering 
component. However, its high- and low-frequency flickering have the same 
origin. They are produced in a two-armed pattern reminiscent of the 
tidally-induced spiral shocks seen in outbursting accretion disks of 
dwarf novae (Steeghs, Harlaftis \& Horne 1997; Baptista et~al. 2002).
These shocks are produced by tidal effects when the disk expands beyond 
the 3:1 resonance radius.  In a dwarf novae these spiral shocks are only
seen during outburst, when the disk becomes hot and large.  In a 
nova-like system, the spiral shocks may be a permanent feature if the
hot disk is large enough for tidal effects to become relevant.
The presence of spiral structures in the steady-light and flickering maps
of UU~Aqr is a likely indication that its accretion disk is large enough
for the tidal pull of the mass-donor star to be relevant for the gas
dynamics in the outer disk regions. This is in line with the detection 
of long-lasting superhumps in this binary (Patterson et~al. 2005) -- 
the tidal influence that leads to spiral density waves may also induce
elliptical orbits in the outer disk regions. Because the spiral arms 
account for a small fraction of the steady-light and are not related to
the broad and brighter steady emission centered in the disk, one might 
conclude that they are not the dominant source of viscous dissipation 
and angular momentum removal in the UU~Aqr accretion disk.

Why do the asymmetric arcs flicker? We discuss three possibilities.
One may consider that flickering is a consequence of unsteady 
dissipation of energy from a clumpy gas stream as it hits the two-armed
spiral density wave in the disk (i.e., a mass transfer origin for the 
flickering, as proposed by Warner \& Nather 1971).
In this case the clumpy gas stream should also lead to detectable 
flickering at the location of the bright spot, where it first hits the
accretion disk rim before reaching the spiral arms.  However, while the
bright spot is a significant light source in the long-term changes map
(indicating that the stream-disk impact occurs at $R\simeq 0.6\,R_{L1}$,
farther out in the disk than the observed radial position of the spiral
arms) it gives no contribution to the flickering. We may conclude that
there is no evidence of clumpiness in the infalling gas stream and,
therefore, we shall exclude this as a viable explanation. 
A second possibility is to consider that flickering arises from
reprocessing at tidally-induced and vertically-thickened disk regions
of unsteady irradiation from the boundary layer (a boundary layer/WD
flickering, Bruch 1992). The problem with this explanation is similar to 
that of the previous one: it would be hard to explain why we do not see 
optical flickering directly from the innermost disk regions.
The third and most promising possibility is to assume that the observed 
flickering is the consequence of turbulence generated by the shock of 
disk gas as it passes through the tidally-induced spiral density waves 
(i.e., a local origin for the flickering).
Further numerical simulations of spiral shocks in accretion disks would
be useful to verify the turbulent nature of the after-shock gas and to
test if such turbulence could generate the observed power-law dependency
[$P(f) \propto f^{-1.5}$] of the resulting energy dissipation fluctuations.

Assuming that the disk gas moves in Keplerian orbits around the $0.67\;
M_\odot$ WD (BSC), the locations of the asymmetric arcs corresponds to 
Doppler velocities of $900-1050 \;km\,s^{-1}$ and $700-800\;km\,s^{-1}$, 
respectively for arcs 1 and 2. Given that the after-shock gas is expected
to have sub-Keplerian velocities ($\simeq 15-40$ per cent lower, see 
Steeghs \& Stehle 1999; Baptista et~al. 2005), these structures may appear 
as arcs of enhanced emission at velocities $\simeq 450-650\;km\,s^{-1}$ 
in the upper left (arc 2) and lower right (arc 1) quadrant on a Doppler 
tomogram. Kaitchuck et~al. (1998) and Hoard et~al. (1998) report that
UU~Aqr disk line-emission is largely asymmetric. Their tomograms show 
regions of enhanced emission which may be interpreted as arising from a
two-armed spiral pattern in the disk. The asymmetry corresponding to arc 
1 is clearly seen (e.g., see Fig.\,6 of Hoard et~al. 1998 and Fig.\,15 of
Kaitchuck et~al. 1998), while that related to arc 2 may be blended with 
and hidden by the emission from the gas stream and bright spot impact site. 
This latter effect may help to explain the large mass ratio ($q=0.86$) 
inferred by Kaitchuck et~al. (1998) by fitting the gas stream trajectory
to the asymmetry in the upper left quadrant of their Doppler tomograms.

\section{Summary}

Our investigation of the sources of variability in UU~Aqr indicates that
the long-term changes giving rise to the 'low' and 'high' brightness 
states can be accounted for by changes in mass transfer rate of 20-50
per cent. A high S/N steady-light light curve reveal the presence of
long-lasting (at least over a 4 yr period of the observations) 
asymmetric arcs in the accretion disk aside of the broad brightness
distributions of a roughly steady-state disk. The arcs are interpreted 
as the consequence of tidally-induced spiral shocks in an extended and 
hot accretion disk. The spiral arms account for 25 per cent of the 
steady-light flux and are the dominant source of flickering, both at 
low- and high-frequencies. They are more tightly wound than the spiral 
shocks found in the outbursting dwarf nova IP~Peg. The observed 
flickering shows a power spectrum density with an $f^{-1.5}$ 
power-law dependency, and is best explained as resulting from turbulence 
generated after the collision of disk gas with the density-enhanced 
spiral wave in the accretion disk.
There is no evidence of flickering originating in the bright spot at
disk rim or in the innermost disk regions around the WD.

\acknowledgements

This work was partially supported by CNPq/Brazil through research grant
62.0053/01-1-PADCT III/Milenio.
RB acknowledge financial support from CNPq/Brazil through grants no. 
300\,354/96-7 and 200\,942/2005-0.

\clearpage

%%%%%%%%%% TABLES
%
%%%%%%%%%%%%%%%%%%%%%%%%%%%%%%%%%%%%%%%%%%%%% TABLE 1
% \tableline\tableline
%  HJD start & $N_p$ & $\Delta\,t$ & E & phase & Quality &
%  HJD start & $N_p$ & $\Delta\,t$ & E & phase & Quality \\ [-0.5ex]
% (2.450.000+) && (s) & (cycle) & range && 
% (2.450.000+) && (s) & (cycle) & range & \\ 
% \tableline
\begin{deluxetable}{crcccccrcccc}
\tabletypesize{\scriptsize}
\tablecaption{Journal of the observations\label{tab1}}
\tablewidth{0pt}
\tablehead{
\colhead{HJD start} & \colhead{$N_p$} & \colhead{$\Delta\,t$} & 
\colhead{E \tablenotemark{a}} & \colhead{phase} & \colhead{Quality 
\tablenotemark{b}} & 
\colhead{HJD start} & \colhead{$N_p$} & \colhead{$\Delta\,t$} & 
\colhead{E \tablenotemark{a}} & \colhead{phase} & \colhead{Quality 
\tablenotemark{b}} \\
\colhead{(2.450.000+)} && \colhead{(s)} & \colhead{(cycle)} & 
\colhead{range} &&
\colhead{(2.450.000+)} && \colhead{(s)} & \colhead{(cycle)} & 
\colhead{range}
}
\startdata
 1047.6259 &  368 & 10 & (28734) & $+0.24,+0.50$ & A & 
 1782.5171 &  999 & 10 & 33227 & $-0.22,+0.48$ & B \\
 1047.6683 & 1446 & 10 & 28735 & $-0.50,+0.50$ & A &  
 1782.6490 &  433 & 10 & (33228) & $-0.42,-0.10$ & C \\ 
 1050.4957 &  506 & 10 & 28752 & $-0.22,+0.15$ & A &  
 1787.6155 &  347 & 10 & 33258 & $-0.05,+0.25$ & C \\ 
 1050.6815 &  339 & 10 & 28753 & $-0.08,+0.17$ & A &  
 2136.5369 &  649 & 10 & 35391 & $-0.03,+0.45$ & B/C \\ 
 1052.4647 & 1179 &  7 & 28764 & $-0.18,+0.41$ & B &
 2136.6232 &  538 & 10 & (35392) & $-0.50,+0.50$ & B/C \\ 
 1052.5936 & 1322 &  7 & 28765 & $-0.39,+0.43$ & B &  
 2137.4796 & 1079 & 10 & 35397 & $-0.27,+0.50$ & B \\ 
 1053.5675 & 1397 &  8 & 28771 & $-0.44,+0.35$ & B/C &  
 2137.6048 & 1413 & 10 & 35398 & $-0.50,+0.50$ & B \\ 
 1053.7460 &  201 & 20 & (28772) & $-0.35,-0.06$ & B/C &  
 2138.4690 &  676 & 15 & 35403 & $-0.22,+0.50$ & B \\ 
 1054.6388 &  169 & 15 & (28777) & $+0.11,+0.29$ & C &  
 2138.5863 &  756 & 15 & 35404 & $-0.50,+0.50$ & B \\ 
 1055.5546 &  265 & 10 & (28783) & $-0.29,-0.06$ & C &  
 2138.7499 &  550 & 15 & 35405 & $-0.50,+0.09$ & B \\ 
 1403.4937 &  898 & 10 & 30910 & $-0.27,+0.37$ & B &  
 2139.4636 &  590 & 15 & 35409 & $-0.14,+0.50$ & B \\
 1403.6550 &  885 & 10 & 30911 & $-0.28,+0.39$ & B & 
 2139.5678 &  886 & 15 & 35410 & $-0.50,+0.50$ & A \\
 1404.4772 & 1068 & 10 & 30916 & $-0.26,+0.50$ & B & 
 2139.7313 &  604 & 15 & 35411 & $-0.50,+0.14$ & A \\
 1404.6009 & 1242 & 10 & 30917 & $-0.50,+0.50$ & B &
 2140.6193 &  380 & 15 & 35416 & $-0.07,+0.33$ & C \\ 
 1404.7644 &  563 & 10 & (30918) & $-0.50,-0.10$ & B &
 2141.5954 &  192 & 15 & 35422 & $-0.10,+0.10$ & B/C \\ 
&&&&&& 2141.7556 & 346 & 15 & 35423 & $-0.13,+0.12$ & B/C \\
[0.5ex]
\enddata
\tablenotetext{a}{ with respect to the ephemeris of eq.~(\ref{efem}).}
\tablenotetext{b}{ sky conditions: A= photometric (main comparison 
stable), B= good (some sky variations), C= poor (large variations 
and/or clouds).} 
\end{deluxetable}
%%%%%%%%%%%%%%%%%%%%%%%%%%%%%%%%%%%%%%%%%%%%%%%%

\clearpage

%%%%%%%%%%%%%%%%%%%%%%%%%%%%%%%%%%%%%%%%% FIG.1
\begin{figure}
%\center
%\includegraphics[scale=0.6,angle=270]{f1.eps}
\caption{Light curves of UU~Aqr (lower panel) and of a comparison star 
  (upper panel) superimposed in phase. The data of the 'high' and 'low'
  brightness states are plotted with black and grey symbols, 
  respectively. (SEE ATTACHED JPG FILE) }
\label{fig1}
\end{figure}
%%%%%%%%%%%%%%%%%%%%%%%%%%%%%%%%%%%%%%%%%%

%\clearpage

%%%%%%%%%%%%%%%%%%%%%%%%%%%%%%%%%%%%%%%%% FIG.2
\begin{figure}
\includegraphics[angle=270,scale=0.6]{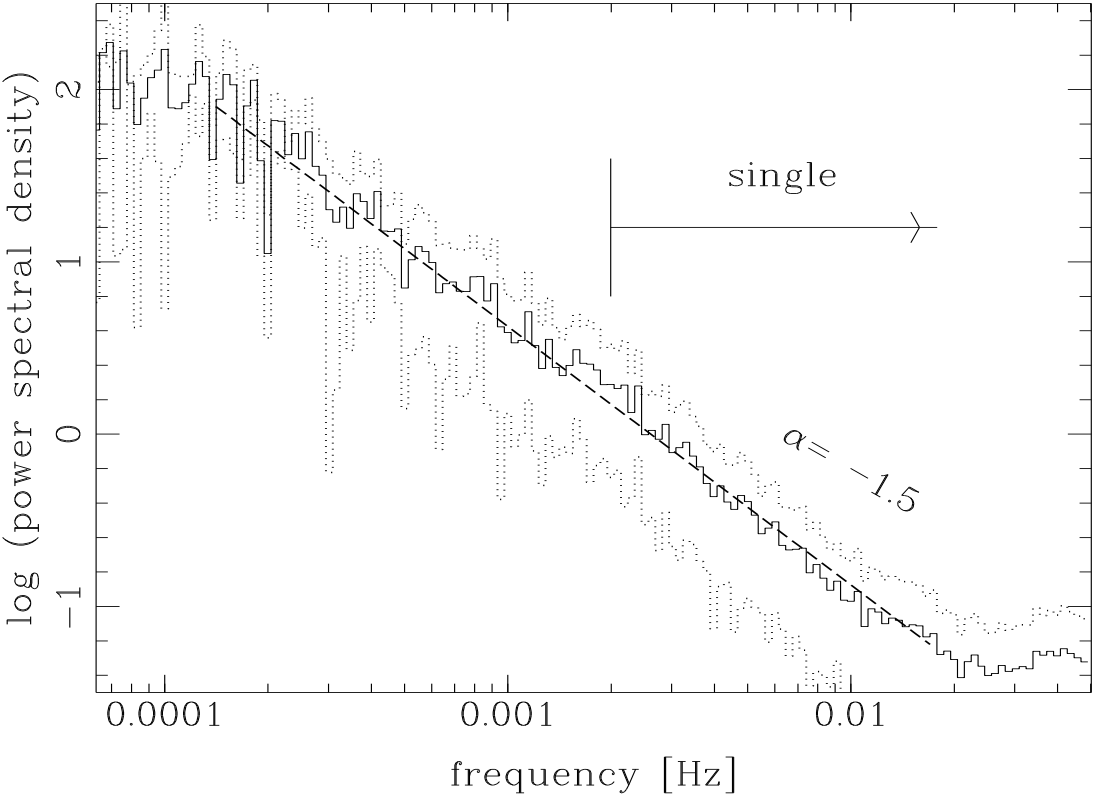}
\caption{Average power density spectrum. Dotted lines show the $1\sigma$ 
  limits on the average power. The best-fit power-law $P(f) \propto 
  f^{-1.5}$ is shown as a dashed line. A vertical tick marks the
  low-frequency cutoff of the filtering process applied to derive the
  'single' scatter curve. }
\label{fig2}
\end{figure}
%%%%%%%%%%%%%%%%%%%%%%%%%%%%%%%%%%%%%%%%%%

\clearpage

%%%%%%%%%%%%%%%%%%%%%%%%%%%%%%%%%%%%%%%%% FIG.3
\begin{figure*}
\center
\includegraphics[bb= 1cm 0.5cm 19.5cm 24.5cm,angle=270,scale=0.7]{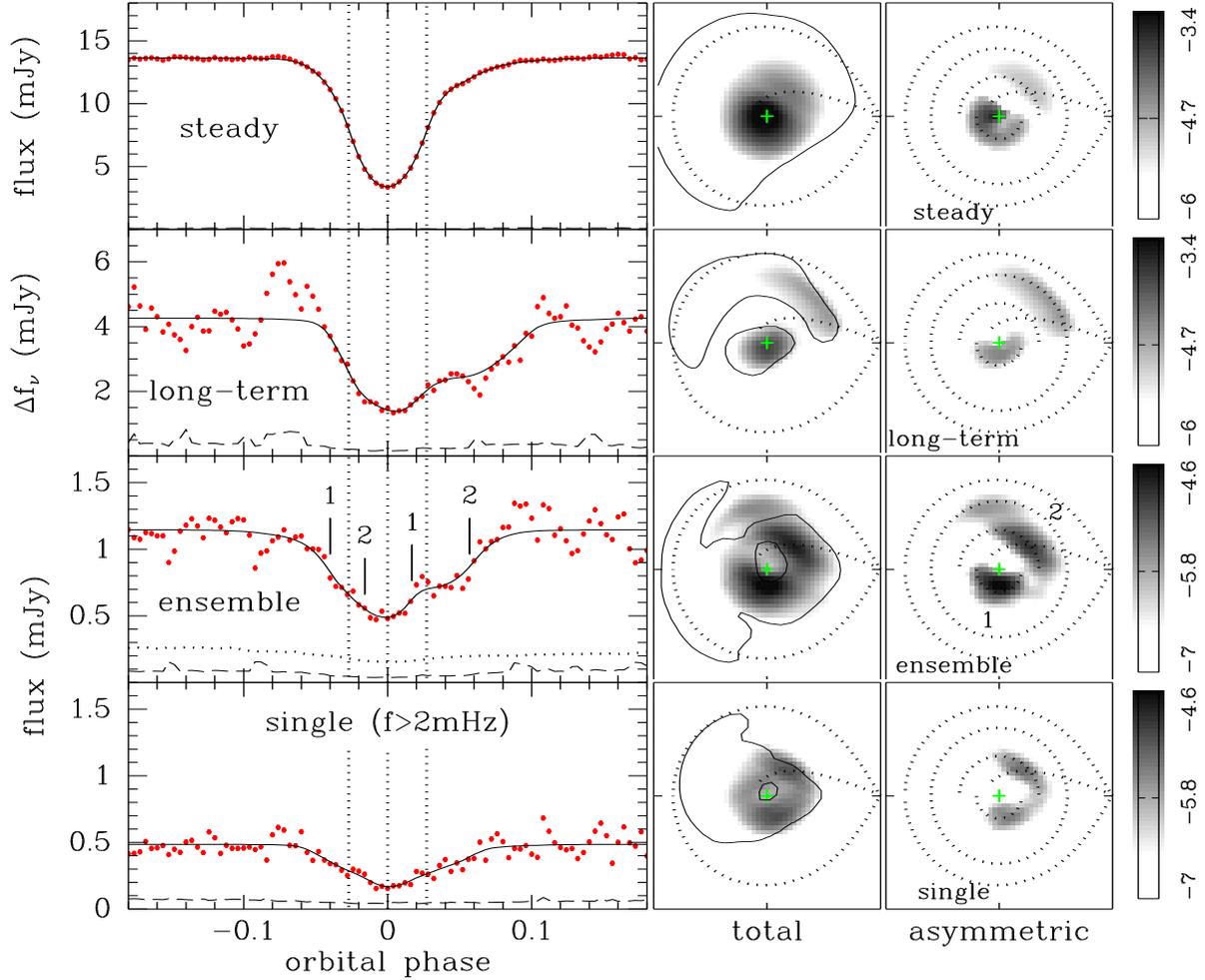}
\caption{Left-hand panels: data (dots) and model (solid lines) light 
 curves of UU Aqr. From top to bottom: the steady-light component, the 
 curve of the long-term changes (secular), the 'ensemble' and the 
 'single' flickering curves.  Vertical dotted lines mark the 
 ingress/egress and mid-eclipse phases of the white dwarf. A dashed 
 curve depicts the error bars in each case. A dotted curve in the 
 'ensemble' panel shows the average contribution of the Poisson noise
 to the scatter at each phase. Vertical ticks mark the ingress/egress 
 phases of the two asymmetric structures seen in the eclipse maps. 
 Middle panels: corresponding eclipse maps in a logarithmic grey-scale.
 A cross marks the center of the disk and a dotted line shows the 
 primary Roche lobe. A solid contour line is overploted on each eclipse
 map to indicate the 3-$\sigma$ confidence level region. Right-hand 
 panels: asymmetric component of the eclipse maps in a logarithmic 
 grey-scale. Additional dotted lines show the ballistic stream 
 trajectory, a reference circle of radius $0.6\;R_{L1}$ and semi-circles
 of radii 0.2 and $0.32\;R_{L1}$. Labels in the 'ensemble' panel depict
 the location of the asymmetric sources 1 and 2. The secondary is to the 
 right of each panel and the stars rotate counter-clockwise. Scale bars 
 in the right side depict the log(intensity) scale in each case. }
\label{fig3}
\end{figure*}
%%%%%%%%%%%%%%%%%%%%%%%%%%%%%%%%%%%%%%%%%%

\clearpage

%%%%%%%%%%%%%%%%%%%%%%%%%%%%%%%%%%%%%%%%% FIG.4
\begin{figure}
\center
\includegraphics[scale=0.5,angle=-90]{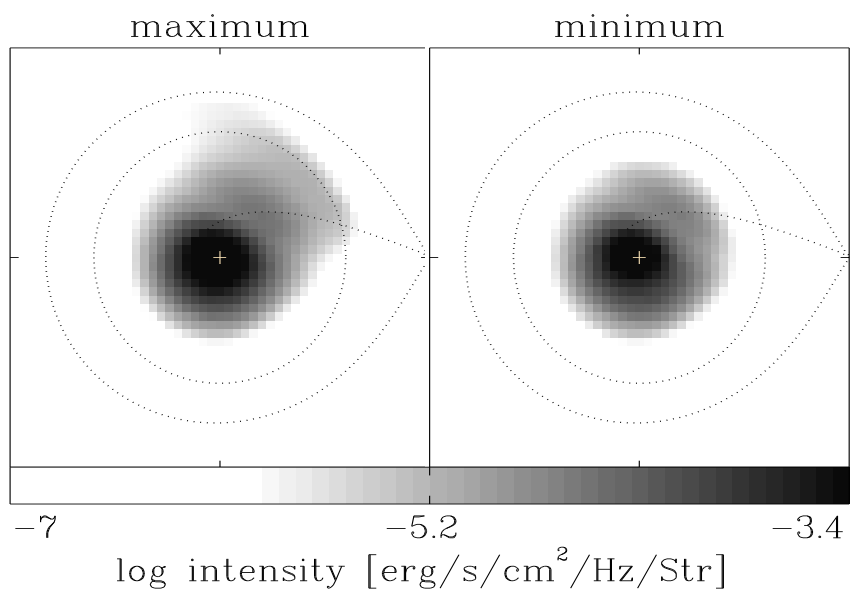}
\caption{Maps of maximum (left) and minimum (right) brightness, obtained 
  from the maps of the steady-light and the long-term changes. The 
  notation is similar to that of Fig.~\ref{fig3}.}
\label{fig4}
\end{figure}
%%%%%%%%%%%%%%%%%%%%%%%%%%%%%%%%%%%%%%%%%%

\clearpage

%%%%%%%%%%%%%%%%%%%%%%%%%%%%%%%%%%%%%%%%% FIG.5
\begin{figure}
\center
\includegraphics[scale=0.5,angle=-90]{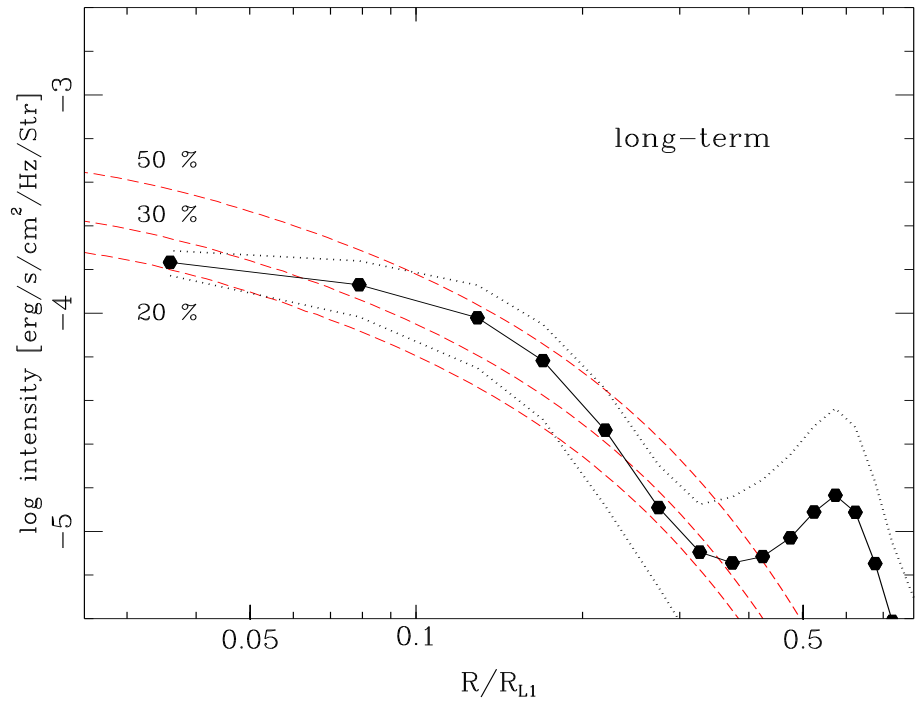}
\caption{Average radial intensity distribution of the long-term changes
  map for an assumed distance of 200~pc (BSH). Dotted lines show the 
  1-$\sigma$ limit on the average intensity. Dashed lines depict the 
  radial intensity distributions resulting from the difference between
  steady-state disk models with mass accretion rates differing by 20, 30 
  and 50 per cent. The reference steady-state opaque disk for these 
  calculations has \.{M}$= 10^{-9.1}\,M_\odot\;yr^{-1}$. }
\label{fig5}
\end{figure}
%%%%%%%%%%%%%%%%%%%%%%%%%%%%%%%%%%%%%%%%%%

\clearpage

%%%%%%%%%%%%%%%%%%%%%%%%%%%%%%%%%%%%%%%%% FIG.6
\begin{figure}
\center
\includegraphics[scale=0.5]{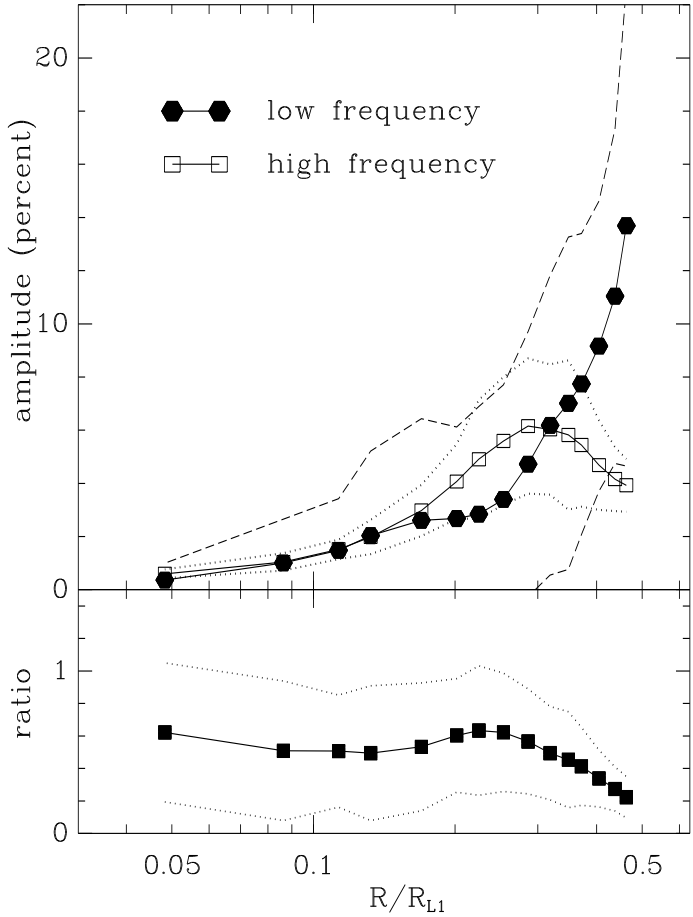}
\caption{Top: the radial run of the relative amplitude of the low- and 
  high-frequency flickering components. Dashed and dotted lines show the
  1-$\sigma$ limits on the average amplitude, respectively for the low-
  and high-frequency flickering. Bottom: the radial run of the fractional
  contribution of the high-frequencies to the total flickering. Dotted 
  lines show the 1-$\sigma$ limits on the distribution. }
\label{fig6}
\end{figure}
%%%%%%%%%%%%%%%%%%%%%%%%%%%%%%%%%%%%%%%%%%

%\bibliographystyle{apj}

\end{document}